\documentclass[conference, 10pt, 1in]{IEEEtran}
\usepackage[font=small,labelfont=bf]{caption}
\usepackage{graphicx, balance}
\usepackage{caption,subcaption}
\usepackage{linguex}
\usepackage[linguistics,edges]{forest}
\usepackage{enumitem}
\usepackage{booktabs}
\usepackage{multirow, makecell}
\usepackage[cmex10]{amsmath}
\usepackage{algorithmic}
\usepackage{url}
\usepackage{csquotes} 
\usepackage{arydshln} 
\usepackage{color, xcolor, soul}
\usepackage{amssymb}
\usepackage{makecell}
\usepackage{cite}
\usepackage{etoolbox}
\usepackage{threeparttable}
\usepackage{hyperref}
\usepackage{pifont}
\usepackage{tikz}
\usetikzlibrary{matrix}
\usepackage{pifont} 
\usepackage{array}
\usepackage{booktabs}
\usepackage{multirow}
\usepackage{tabularx}

\usepackage{rotating}
\usepackage{booktabs}

\newcolumntype{C}[1]{>{\centering\arraybackslash}p{#1}}

\definecolor{StartStopColor}{HTML}{FC427B} 
\definecolor{InputOutputColor}{HTML}{00FFFF} 
\definecolor{ProcessColor}{HTML}{50C878} 
\definecolor{DecisionColor}{HTML}{A7257D}
\definecolor{textcolor}{HTML}{000000}
\definecolor{processtextcolor}{HTML}{000000}

\usepackage{tikz}
\usetikzlibrary{shapes.geometric, arrows}
\tikzstyle{startstop} = [rectangle, rounded corners, minimum width=3cm, minimum height=1cm, text centered, draw=black, text=textcolor, fill=StartStopColor!80]
\tikzstyle{io} = [trapezium, trapezium left angle=70, trapezium right angle=110, text width=5cm, minimum height=1cm, text centered, draw=black, fill=InputOutputColor!50, text=textcolor]
\tikzstyle{process} = [rectangle, text width=8cm, minimum height=1cm, text centered, draw=black, fill=ProcessColor!80, text=textcolor]
\tikzstyle{decision} = [diamond, minimum width=3cm, minimum height=1cm, text centered, draw=black, fill=DecisionColor]
\tikzstyle{arrow} = [thick,->,>=stealth]

\tikzstyle{io_1} = [trapezium, trapezium left angle=70, trapezium right angle=110, text width=6cm, minimum height=1cm, text centered, draw=black, fill=InputOutputColor, text=textcolor]
\tikzstyle{process_1} = [rectangle, text width=5cm, minimum height=1cm, text centered, draw=black, fill=ProcessColor, text=textcolor]

\title{Phish-Blitz: Advancing Phishing Detection with Comprehensive Webpage Resource Collection and Visual Integrity Preservation}

\author{
    \IEEEauthorblockN{
        Duddu Hriday\textsuperscript{*},
        Aditya Kulkarni\textsuperscript{*},
        Vivek Balachandran\textsuperscript{\#}, and 
        Tamal Das\textsuperscript{*}
    }

    \IEEEauthorblockA{
        \textsuperscript{*}\emph{Indian Institute of Technology, Dharwad, India} \\
        \textsuperscript{\#}\emph{Singapore Institute of Technology, Singapore} \\
        \textsuperscript{*}\{210010016, aditya.kulkarni, tamal\}@iitdh.ac.in, \textsuperscript{\#}vivek.b@singaporetech.edu.sg}
}

\begin{document}

\maketitle

\definecolor{StartStopColor}{HTML}{FFFFFF} 
\definecolor{InputOutputColor}{HTML}{FFFFFF} 
\definecolor{ProcessColor}{HTML}{FFFFFF} 
\definecolor{DecisionColor}{HTML}{FFFFFF} 
\definecolor{textcolor}{HTML}{000000} 
\definecolor{processtextcolor}{HTML}{FFFFFF} 

\tikzstyle{startstop} = [rectangle, rounded corners, minimum width=1.75cm, minimum height=1cm, text centered, draw=black, text=textcolor, fill=StartStopColor]
\tikzstyle{io} = [trapezium, trapezium left angle=70, trapezium right angle=110, text width=4.5cm, minimum height=1cm, text centered, draw=black, fill=InputOutputColor, text=textcolor]
\tikzstyle{process} = [rectangle, text width=5.5cm, minimum height=1cm, text centered, draw=black, fill=ProcessColor, text=textcolor]
\tikzstyle{decision} = [diamond, minimum width=3cm, minimum height=1cm, text centered, draw=black, fill=DecisionColor]
\tikzstyle{arrow} = [thick,->,>=stealth]

\tikzstyle{io_1} = [trapezium, trapezium left angle=70, trapezium right angle=110, text width=6cm, minimum height=1cm, text centered, draw=black, fill=InputOutputColor, text=textcolor]
\tikzstyle{process_1} = [rectangle, text width=5cm, minimum height=1cm, text centered, draw=black, fill=ProcessColor, text=textcolor]

\begin{abstract}
Phishing attacks are increasingly prevalent, with adversaries creating deceptive webpages to steal sensitive information. Despite advancements in machine learning and deep learning for phishing detection, attackers constantly develop new tactics to bypass detection models. As a result, phishing webpages continue to reach users, particularly those unable to recognize phishing indicators. To improve detection accuracy, models must be trained on large datasets containing both phishing and legitimate webpages, including URLs, webpage content, screenshots, and logos. However, existing tools struggle to collect the required resources, especially given the short lifespan of phishing webpages, limiting dataset comprehensiveness. 

In response, we introduce Phish-Blitz, a tool that downloads phishing and legitimate webpages along with their associated resources, such as screenshots. Unlike existing tools, Phish-Blitz captures live webpage screenshots and updates resource file paths to maintain the original visual integrity of the webpage. We provide a dataset containing 8,809 legitimate and 5,000 phishing webpages, including all associated resources. Our dataset and tool are publicly available on GitHub, contributing to the research community by offering a more complete dataset for phishing detection.
\end{abstract}


\section{Introduction}
\label{sec:Introduction}
As of April 2024, there were 5.44 billion Internet users worldwide, which amounted to 67.1 percent of the global population \cite{2024internet}. This widespread reliance on the Internet has made it a prime target for cyberattacks. In 2023, the average cost of a data breach worldwide increased to USD 4.45 million, up by USD 100,000 from 2022. This marks a 2.3\% rise from the 2022 average cost of USD 4.35 million \cite{kuzior2024cybersecurity}. To combat these threats, cybersecurity measures are essential. One of the most common types of cyber attacks is phishing. Phishing refers to a set of cyberattacks where deception and disguising techniques are employed by an attacker to gain access to the
victim’s sensitive information on the Internet \cite{varshney2024anti}. Different types of phishing attacks exist, which involve sending phishing URLs to the users such as email-phishing, SMS phishing, vishing (voice-based), pharming, pop-up phishing, and many more. Attackers pretend to be legitimate entities such as banks or popular online services, to trick users into entering their credentials or personal information. \par
\begin{figure}[t]
    \centering
    \includegraphics[width=\linewidth]{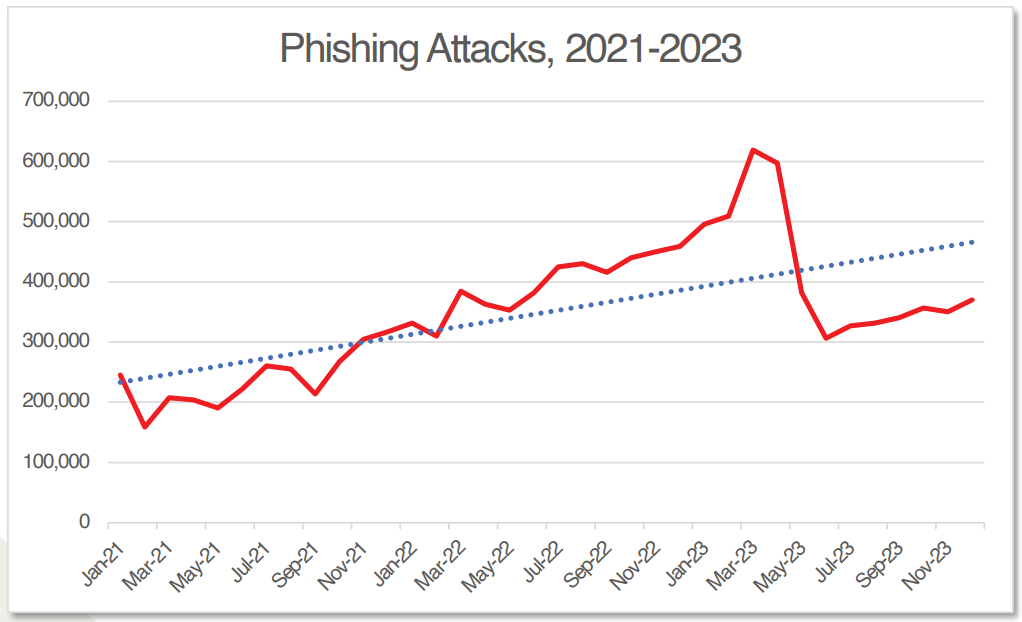} %
    \caption{Phishing Attacks from Jan 2021 to Nov 2023}
    \label{fig:phishing_attack_trend}
\end{figure}

The Anti-Phishing Working Group (APWG) is an international coalition that attempts to eliminate fraud and identity theft caused by phishing and related incidents. APWG has observed almost 5 million phishing attacks over the course of 2023, which is considered a record year. In the first quarter of 2024, APWG observed 963,994 phishing attacks. The trend of phishing attacks as observed by APWG \cite{APWG_REPORT_4_2022} is shown in Figure \ref{fig:phishing_attack_trend}. Although the numbers appear to decrease relatively, they remain sufficiently high to be effective in this field. \par

Phishing is not a recent problem. According to reports, phishing attacks originated back in the 1990s \cite{wikiphish}. Since then there have been many approaches to tackle these attacks. The early phishing detection techniques were a \textit{whitelist} (list of legitimate sites) and a \textit{blacklist} (list of phishing sites) based techniques. However, this technique could not catch phishing sites that were not part of this list. It was followed by heuristic-based approaches to detect phishing sites. This detection technique involved examining URL characteristics such as the domain, primary domain, subdomain, and path domain. These features are analyzed and used to detect phishing sites based on the gathered information. In recent times, most of the phishing detection techniques involve Machine Learning (ML) and Deep Learning (DL) \cite{kulkarni2024phishing}. These approaches use URL-based, content-based, screenshot-based and third-party-based features for phishing webpage detection.  \par



Datasets play a crucial role in phishing webpage detection by providing labelled examples that train and evaluate ML models, enabling the identification of malicious patterns. They help in extracting features and benchmarking different detection techniques, helping in the development of more accurate and robust models. Existing techniques for phishing detection, utilize tools such as \texttt{Selenium WebDriver}\cite{selenium}, \texttt{PyWebCopy} \cite{pywebcopy}, \texttt{BeautifulSoup} \cite{beautifulsoup}, \texttt{requests} \cite{requests}, and \texttt{wget} \cite{wget} among others, to fetch datasets. These tools download the basic HTML page of the website and some of these tools also download the related CSS, JS, and image files along with HTML. However, the dataset used in phishing webpage detection needs to include screenshots too, as the DL methods mainly depend on the screenshot of the web pages. Another crucial aspect is that the HTML file downloaded by these tools contain links (\texttt{script}, \texttt{link}, \texttt{img} tags), that point to the files uploaded on the server of the website i.e., the \texttt{src} and \texttt{href} attributes, contain links of the files uploaded on the server, rather than pointing to the locally downloaded resources. This causes a problem mainly when phishing webpages are downloaded. The study reveals that phishing webpages have a remarkably short active lifespan. 

The sharp decline in active phishing webpages is significant, with a drop from 65\% to 53\% in just the first five minutes, and less than 40\% still active after 24 hours \cite{skula2024phishing}. When the phishing webpage expires, most of the links pointing to the files on the server, become invalid. This will impact the appearance and front-end view of the downloaded webpage. In case of legitimate webpages also, the locally downloaded webpage appearance entirely changes when viewed offline. This is because, when the resources available on the server are down or when the Internet is not available, files like CSS, JS and images does not render on the HTML file.

To address these issues, it is essential to develop a methodology that not only downloads the webpage and its associated resources but also modifies the links to reference these resources locally, ensuring the webpage remains fully functional offline. Additionally, capturing screenshots helps in training phishing detection models using DL.

In this paper, we present a tool -- Phish-Blitz -- that not only downloads the basic HTML page of the website, but also downloads all the associated web resources i.e. images, javascript, and CSS files. Along with downloading these resources, the tool also updates the reference of the CSS, JS, and image files (\texttt{src, href} attributes) to point to the locally downloaded resources. This ensures that the downloaded webpage accurately replicates the appearance and structure of the original website. This helps in analysing the webpage as the downloaded webpage is unaffected, even if the original website is taken down or if the links pointing to the web resources (CSS, JS, images) become invalid. Along with downloading all these resources, the tool also takes screenshots of the downloaded webpage and the original website. Using this tool, we have collected a dataset containing 5000 phishing webpages and 8809 legitimate webpages. We have made these resources public, by uploading them on a public GitHub repository \cite{our_tool}. This dataset can be used effectively for content-based, screenshot-based, and logo-based phishing detection techniques, thus enhancing the robustness and reliability of phishing detection research. \par


The rest of this paper is organized as follows. Section \ref{sec:Related_Work} describes the existing open-source tools for webpage resource collection and describes our contributions. Section \ref{sec:Proposed_Work} describes the working of our tool Phish-Blitz. Section \ref{sec:Experiment_Setup_Results_And_Analysis} describes the internal working of the tool and the resources used to build this tool, the observations made using Phish-Blitz, and the comparative analysis of Phish-Blitz with two other existing tools for web resource collection. Section \ref{sec:Conclusion} presents few concluding remarks and the future scope of Phish-Blitz.

 \begin{table*}[ht]
\caption{URL and Content-based Features}
\label{tab:URL_and_content_based_features}
\begin{tabular}{cll}
\toprule
\textbf{Type} & \textbf{Function Name} & \textbf{Explanation} \\ \midrule 
\multirow{10}{*}{\rotatebox[origin=c]{90}{\textbf{URL}}} & \texttt{domain\_is\_IP} & Checks if the domain part of the URL is an IP address. \\ 
 & \texttt{symbol\_count} & Counts the number of special symbols such as @, -, \~. \\ 
 & \texttt{https} & Checks if the URL uses HTTPS. \\ 
 & \texttt{domain\_len} & Returns the length of the full domain, including subdomain and TLD. \\
 & \texttt{url\_len} & Returns the length of the entire URL. \\ 
 & \texttt{sensitive\_word} & Returns True if any of the sensitive words like ``secure'', ``login'', ``banking'' are present in the URL. \\ 
 & \texttt{tld\_in\_domain} & Checks if any TLD (top-level domain) appears in the domain or subdomain part of the URL. \\ 
 & \texttt{tld\_in\_path} & Checks if any TLD appears in the path, parameters, query, or fragment of the URL. \\ 
 & \texttt{https\_in\_domain} & Checks for the presence of \texttt{https} in the domain part of the URL. \\ 
 & \texttt{abnormal\_url} & Returns True, if the domain and hostname of the URL do not match. \\ \midrule

\multirow{16}{*}{\rotatebox[origin=c]{90}{\textbf{Content}}} & \texttt{len\_html\_tag} & Returns the sum of lengths of \texttt{style}, \texttt{link}, \texttt{form}, \texttt{script} tags, and comments. \\ 

 & \texttt{len\_html} & Returns the length of HTML content as plain text. \\ 
 & \texttt{hidden} & Returns True, if there are hidden \texttt{div}, \texttt{input}, or \texttt{button} tags. \\ 
 & \texttt{internal\_external\_link} & Returns the counts of internal and external links (anchor tag \texttt{href} only). \\ 
 & \texttt{empty\_link} & Counts empty or placeholder links such as \texttt{href=``''}, \texttt{href=``\#''}, \\ && \quad \texttt{href=``\#javascript::void(0)''}, etc. \\ 
 & \texttt{login\_form} & Returns True, if the ``\texttt{name}'' attribute of input tags in a \texttt{form} is either ``password'', ``pass'', ``login'', \\ && \quad or ``signin''. \\ 
 & \texttt{internal\_external\_resource} & Counts the number of internal and external resources (\texttt{link}, \texttt{img}, \texttt{script}, \texttt{noscript} tags). \\ 
 & \texttt{redirect} & Checks if HTML content contains the string ``redirect''. \\ 
 & \texttt{alarm\_window} & Checks if any \texttt{script} tag contains \texttt{alert} or \texttt{window.open} function. \\ 
 & \texttt{title\_domain} & Checks if the domain appears in the title tag of the HTML content. \\ 
 & \texttt{brand\_freq\_domain} & Counts how many well-known brand names appear in the URL. \\ 
 & \texttt{is\_link\_valid} & Checks if the response code is $200$ or not for all links in the HTML. \\ 
 & \texttt{multiple\_https\_check} & Checks if \texttt{https:} is present more than once in all the links in HTML. \\ 
 & \texttt{form\_empty\_action} & Checks if the action attribute of the form tag is empty or \texttt{about:blank}. \\ 
 & \texttt{is\_mail} & Checks the presence of the mail function or \texttt{mailto} attribute. \\ 
 & \texttt{status\_bar\_customization} & Checks if HTML has the onmouseover attribute and uses \texttt{windows.status}. \\ \bottomrule
\end{tabular}
\end{table*}

\section{Related Work and Our Contributions}
\label{sec:Related_Work}

\subsection{Dataset Repositories}
Several open-source repositories provide phishing and legitimate samples, which are valuable for research, development, and testing in the field of cybersecurity. A few notable ones are as follows.

PhishTank \cite{phishtank} is a community-based anti-phishing site that collects and shares information about phishing webpages on the Internet. It was launched in 2006 by David Ulevitch as an offshoot of OpenDNS \cite{wiki}. PhishTank allows users to submit websites they suspect are phishing scams, and other users vote on whether they agree. The site also maintains a database of verified phishing websites. PhishTank provides accurate and timely information to help researchers building security tools. 

OpenPhish \cite{openphish} is a phishing intelligence platform that helps organizations prevent and detect phishing attacks. It has a database of phishing webpages that contains structured and searchable information, as well as metadata that can be used to analyze cyber incidents, search for trends, and train or validate AI applications.

Alexa \cite{alexa_db}, an Amazon-owned company, provides a dataset of the top-ranked websites globally, known as Alexa Top Sites. This dataset includes URLs of legitimate websites, ranked by traffic. It is often used as a benchmark for legitimate web traffic and can be useful for comparing against phishing datasets. Researchers can use this data to identify common features of legitimate sites and improve the accuracy of phishing detection models. However, Alexa is now outdated, and its database is no longer accessible.

Common Crawl \cite{common_crawl} is a non-profit organization that crawls the web and freely provides its data to the public. The dataset includes petabytes of webpage data, including raw HTML, metadata, and text content. Researchers can use Common Crawl to obtain legitimate web content at scale, providing a rich resource for various analyses, including phishing detection. The breadth and depth of Common Crawl's data make it an invaluable tool for web research and machine-learning projects.

\subsection{Webpage Resource Collectors}
There are several existing tools and libraries available that facilitate the downloading of webpage resources. These tools can capture various elements such as HTML files, CSS, JavaScript, images, and other associated resources needed to render a webpage on localhost. Some of the relevant tools are as follows. 

PyWebCopy \cite{pywebcopy} is a Python library for copying full or partial websites locally onto the hard-disk for offline viewing. PyWebCopy scans the specified website and downloads its content, although the files in the local folder is not well-organized. Links to resources such as style-sheets, images, and other pages in the website are automatically remapped to match the local path. Using its extensive configuration, one can define which parts of a website will be copied and how. 

\texttt{scrapio} \cite{kulkarni2023bridging} is a versatile tool designed for downloading essential web resources such as HTML, CSS, JavaScript, images, favicons, and screenshots. It supports the retrieval of both legitimate and phishing webpages, with phishing data sourced from PhishTank \cite{phishtank}. However, PhishTank has recently removed access to legitimate URLs and no longer permits web scraping, presenting challenges for data acquisition from this platform.

\texttt{saveweb2zip} \cite{saveweb2zip} is a website application, that can be used to download web resources like HTML, CSS, JS, favicon and image files. This application also redirects the links in HTML file to point to locally downloaded resources. The application takes a URL as an input and downloads a zip file containing the necessary web resources. But only one URL can be processed at a time.

\subsection{Discussion}
Although these open-source tools can retrieve essential resources such as HTML, CSS, JS, and image files, they do not offer functionality for capturing screenshots of webpages. They also do not link the locally downloaded files to the HTML file. Tools like PhishTank \cite{phishtank}, provide us with the regularly updated phishing URLs dataset along with the screenshot of the webpage, but it restricts the downloading of the web resources. In research, especially related to phishing webpage detection, we require a tool, that does download all the required web resources and also takes screenshots of the webpage. The main motive to do so is that most of the phishing detection techniques in the current era are mainly based on ML and DL models. So, there is a requirement for a tool that aids researchers in looking into both directions (ML and DL).

\subsection{Our Contributions}
Our key contributions in this paper are as follows.
\label{subsec:Contributions}
\begin{enumerate}
    \item We develop a tool Phish-Blitz, that downloads the HTML file of the webpage along with all the required resources i.e. CSS, JS, and image files. It ensures that the webpage remains the same, even if it is viewed after the original webpage is taken down. It also takes the required screenshots, used for comparative analysis.
    \item Phish-Blitz also extract features. It can fetch some URL-based and content-based features, which are well-known to be common in phishing webpages and can be used to distinguish between legitimate and phishing webpages.
    \item Using Phish-Blitz, we have successfully collected a dataset containing 8809 legitimate webpages and 5000 phishing webpages.
    \item The source code of Phish-Blitz is publicly accessible on GitHub \cite{our_tool}.
\end{enumerate}

\begin{figure*}[ht]
    \centering
    \begin{subfigure}{0.49\textwidth}
        \centering
        \includegraphics[width=1.0\textwidth,]{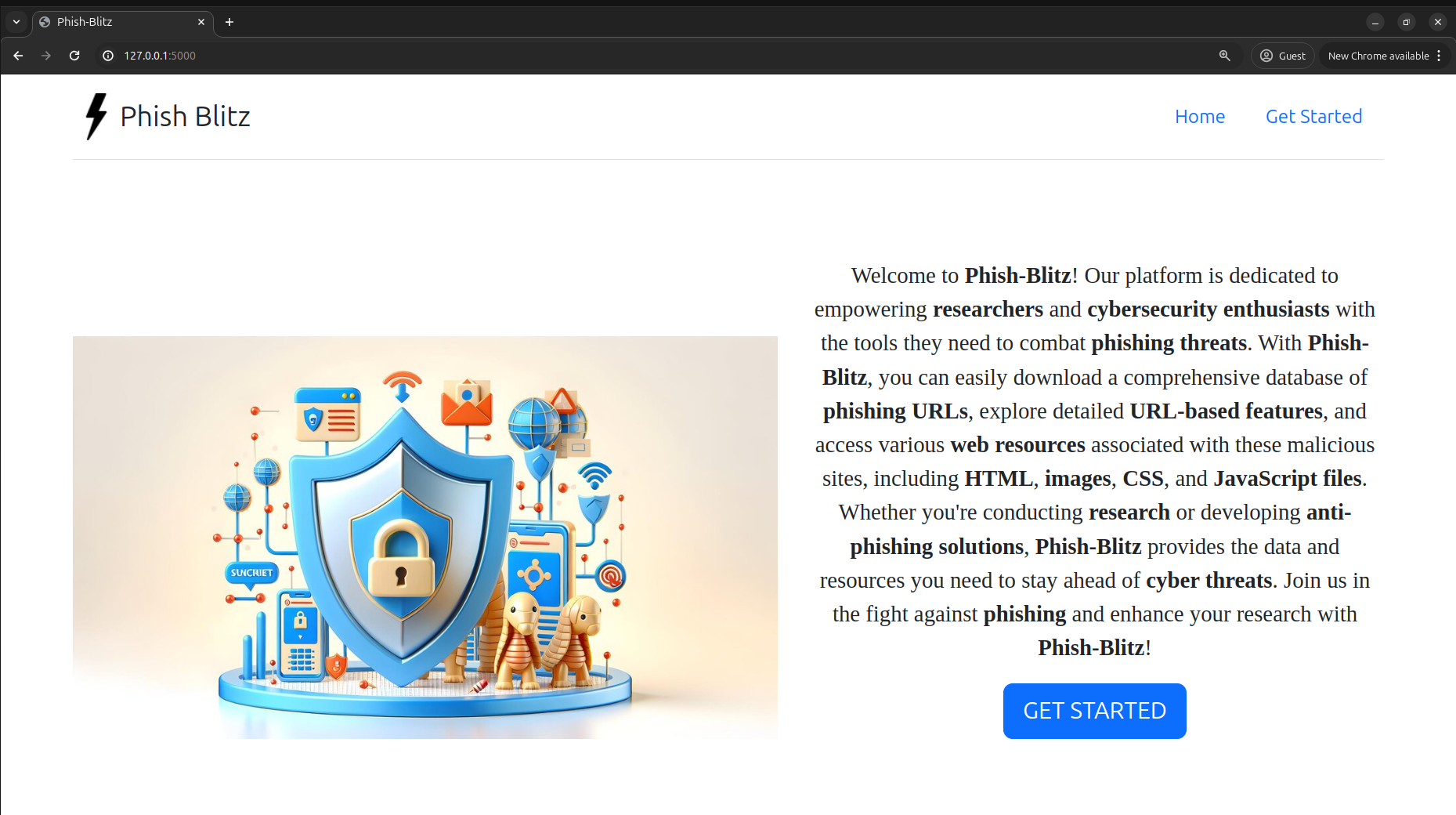}
        \caption{Home Page}
        \label{fig:homepage}
    \end{subfigure}
    \hfill
    \begin{subfigure}{0.49\textwidth}
        \centering
        \includegraphics[width=1.0\textwidth,]{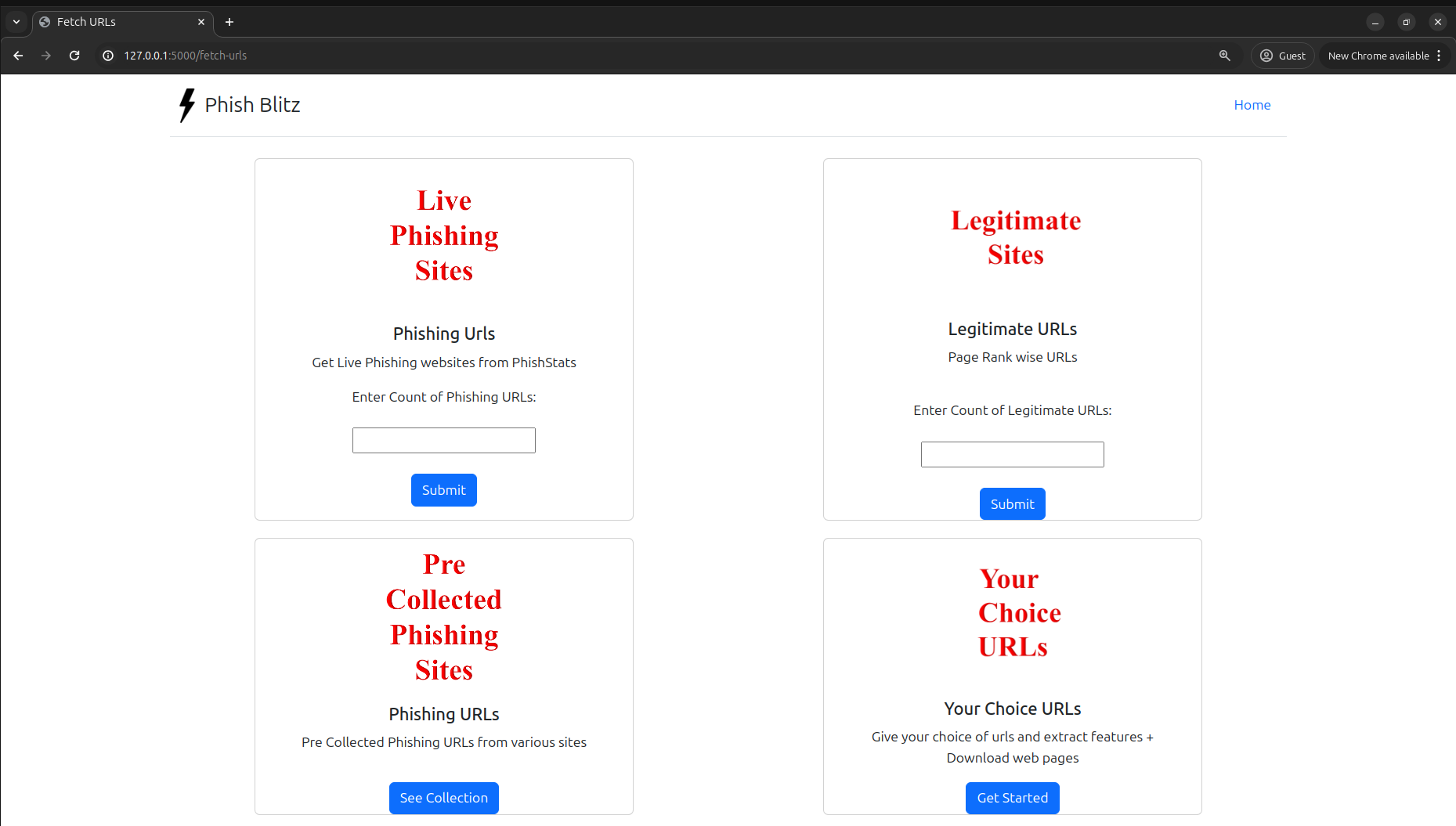}
        \caption{Main Page}
        \label{fig:mainpage}
    \end{subfigure}
    \caption{Phish-Blitz Web Application}
\end{figure*}


\section{Phish-Blitz: Dataset Collector}
\label{sec:Proposed_Work}

In this section, we introduce our tool -- Phish-Blitz -- which downloads HTML, CSS, JS, images and webpage screenshots for both phishing and legitimate webpages. The tool also updates the path of the links of CSS, JS, and images to point to the locally downloaded resources. This step is important for two purposes. 
\begin{enumerate}
    \item The webpage can be viewed offline with all the resources included. If the link path is not changed, then the website looks incomplete when viewed offline because CSS, JS, and images are not loaded.

    \item Phishing webpages are short-lived. So, if downloaded phishing webpage is viewed after its expiry, then the resources may not load at all, as most of the links may not be reachable.
\end{enumerate}



The tool is a web application and it works on localhost. Using localhost, enables the users to run the code on their local machines and also make changes in the code in a way they want. This web application contains different pages with different functionalities. The home page contains an introduction about the tool as shown in Figure~\ref{fig:homepage}. The main page of Phish-Blitz consists of four parts i.e. legitimate URLs, phishing URLs, user-defined URLs and pre-collected database as shown in Figure~\ref{fig:mainpage}.

To fetch the web resources of phishing and legitimate websites, URLs are required. For legitimate URLs, DomCop \cite{top10mwebsites} is used, which provides the top $10$ million legitimate website URLs based on Common Crawl \cite{common_crawl}. The live phishing URLs are collected from PhishStats~\cite{phishstats}. Phish-Blitz allows users to either enter the count of phishing URLs to crawl from PhishStats~\cite{phishstats} or provide a list of URLs of their choice.

After retrieving the URL, the process can continue in two different directions. The first involves extracting URL-based features referenced from ~\cite{Lee2020}. The second direction involves downloading the webpage resources for each of the URLs. In this, Phish-Blitz downloads the landing webpage for the given URL, and checks for the associated resources by parsing through various HTML tags within the downloaded landing webpage, including \texttt{link}, \texttt{script}, \texttt{img} tags to download CSS, Javascript, and image files respectively. In addition to these webpage resources, Phish-Blitz also captures screenshots of the original webpage and the locally downloaded webpage. To ascertain the successful downloading of webpages, a comparative analysis is performed that checks the visual similarity between the two webpage screenshots. A metric known as histogram correlation~\cite{histcorr} is employed, which evaluates the similarity between the two histograms. This method is commonly applied in image processing, particularly in object recognition, image retrieval, and computer vision, where the objective is to determine how similar two datasets are. Once the webpage resources are downloaded, the content-based features are extracted, which are also referenced from \cite{Lee2020}. Table~\ref{tab:URL_and_content_based_features} describes the URL and content-based features.

\begin{table}[t]
    \centering
    \caption{\texttt{wget} options to download webpage resources}
    \label{tab:parameters}
    \begin{tabular}{ll}
    \toprule
    \textbf{Options} & \textbf{Description} \\ \midrule
    \texttt{--mirror} & Captures an entire website with all its \\ & \quad associated resources \\
    \texttt{--adjust-extension} & Adds appropriate extensions to downloaded \\ & \quad files \\
    \texttt{--page-requisites} & Ensures all necessary files for displaying a \\ & \quad webpage are downloaded \\
    \texttt{--no-parent} & Prevents \texttt{wget} from following links outside \\ & \quad of specific site/directory \\
    \texttt{--U} & Used to specify \texttt{user\_agent} (a string that \\ & \quad a web browser or other client software \\ & \quad (like \texttt{wget}), sends to a web browser to \\ & \quad identify itself. \\ \bottomrule
    \end{tabular}
\end{table}

\section{Tool Setup, Observations and Analysis}
\label{sec:Experiment_Setup_Results_And_Analysis}
This section describes the setup of Phish-Blitz tool to download webpage resources for given URLs, analyse the efficiency of Phish-Blitz in downloading the webpage resources and compare it with some existing tools and libraries. 
\subsection{Tool Setup}
\label{sec:Experimental_Setup}
Phish-Blitz fetches the basic HTML content of the landing webpage for the given URL using \texttt{wget}~\cite{wget} and stores the file as \texttt{index.html}, in most cases. \texttt{wget} is a program launched through the command line that serves as a free utility for downloading files from the web. The command line argument used to fetch the HTML file is as follows.

\texttt{wget --mirror --adjust-extension --page-requisites --no-parent --U user\_agent URL}

The options used in the command customize the behaviour of the \texttt{wget} to fit our specific needs. Table~\ref{tab:parameters} describes each of these options.


The set of user agents used with \texttt{wget} command to download the webpage resources are as follows: 
\begin{itemize}
    \item Mozilla/5.0 (Windows NT 10.0; Win64; x64) AppleWebKit/537.36 (KHTML, like Gecko) Chrome/58.0.3029.110 Safari/537.3
    \item Mozilla/5.0 (Windows NT 10.0; Win64; x64; rv:89.0) Gecko/20100101 Firefox/89.0 
    \item Mozilla/5.0 (Windows NT 10.0; Win64; x64) AppleWebKit/537.36 (KHTML, like Gecko) Chrome/58.0.3029.110 Safari/537.3
    \item Mozilla/5.0 (Macintosh; Intel Mac OS X 10\_15\_7) AppleWebKit/537.36 (KHTML, like Gecko) Chrome/91.0.4472.114 Safari/537.36 
\end{itemize}

For every URL, a user agent is randomly chosen from this array, so that the website does not block our IP address, which may result in failure to download webpages.

After downloading the HTML file, the tool crawls through this HTML file and fetches the links of all the required resources, namely CSS, JS, and image files. A Python library, \texttt{BeautifulSoup} \cite{beautifulsoup} is used to crawl through the HTML file. \texttt{BeautifulSoup} does two main works in this case.
\begin{enumerate}
    \item Crawl through the HTML file and fetch the links of all the CSS, JS, and image files.
    \item Change the links to CSS, JS, and image files to point to the locally downloaded web resources.
\end{enumerate}

To download the web resources, a Python library called \texttt{requests} \cite{requests} is used. The \texttt{requests} library is a standard way of making HTTP requests in Python. Phish-Blitz uses \texttt{requests} to download all the valid CSS, JS, and image files from the HTML file. Downloading of the web resources is parallelized to enhance efficiency and reduce overall download time, allowing multiple requests to be handled simultaneously, thereby optimizing resource collection and ensuring a more responsive and effective data-gathering process.

After the tool fetches all the web resources, it checks the visual similarity of the downloaded HTML file with the original webpage, in order to ensure that Phish-Blitz is properly downloading all the necessary resources and redirecting the links. So, the tool takes screenshots of the website and our downloaded webpage (HTML file). To do so, a Python library called \texttt{Selenium}~\cite{selenium} is used along with \texttt{ChromeDriver}~\cite{chromedriver} in headless mode making it possible to automate browsing and taking the required screenshots. The browser window size is set to the full height of the webpage to capture the entire content.

\begin{figure}[t]
    \centering
    \includegraphics[width=0.8\linewidth]{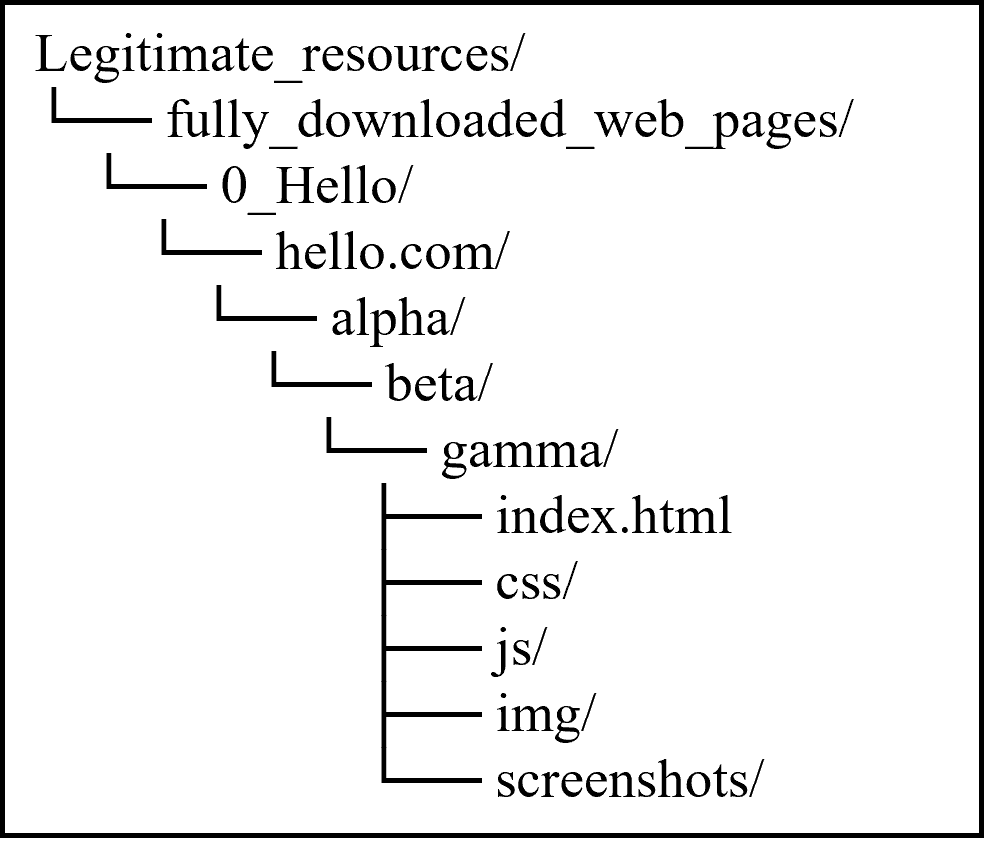}
    \caption{Directory Structure of an example URL}
    \label{fig:dir_struct}
\end{figure}

The directory structure of the downloaded webpage depends on the URL. The outermost directory is named \texttt{legitimate\_resources} in case of legitimate URLs and for user input URLs (your choice URLs). For phishing URLs, it is named \texttt{phishing\_urls}. Inside this outermost directory, there is one directory for every URL. Let us understand the directory structure with an example. Consider a dummy URL \texttt{https://hello.com/alpha/beta/gamma}, then the directory structure will be as shown in Figure \ref{fig:dir_struct}.

 This directory structure for the HTML file is followed by \texttt{wget}. In the directory containing the HTML file, four more directories are created. These folders are created to place CSS, JS, image files, and screenshots separately.

\subsection{Observations}
\label{sec:Experiment_Results}
We have gathered a dataset of web resources from 8809 legitimate websites and 5000 phishing webpages where the URLs have been taken from various resources. This dataset is made publicly available for the users. Apart from that, for our observational study, we have collected a simple collection of urls of 1000 legitimate sites (Page rank based) and 1000 phishing sites (from PhishStats). Out of these URLs, the tool is able to extract web resources for a total of 818 legitimate sites and 597 phishing sites. For our study, the \texttt{requests} library is also used to check how many of the collected URLs are working. To check the success of Phish-Blitz, it is compared with \texttt{requests} library of Python. \texttt{requests} library return a response code 200 for the URLs whose server is responding. It is observed that 811 (out of 1000) legitimate URLs and 592 (out of 1000) phishing URLs are responding to \texttt{requests} library.
From this count, it is observed that :
\begin{itemize}
    \item Phish-Blitz outperforms the \texttt{requests} library as it can achieve a success rate of $81.8\%$ whereas \texttt{requests} has a  success rate of $81.1\%$ with legitimate URLs.
    \item Even in case of phishing URL, Phish-Blitz fetches 597 URLs where as \texttt{requests} is able to fetch 592 URLs.
\end{itemize}

Although Phish-Blitz has a high success rate for legitimate URLs, we have observed that the count of phishing resources is comparatively low. This can be due to any of the following reasons: 
\begin{itemize}
    \item Phishing webpages are typically short-lived because they are detected and taken down rapidly by cybersecurity teams, law enforcement, or hosting providers. This leads to many phishing URLs becoming inactive shortly after they are reported.
    \item The time it takes for a phishing URL to be added to databases like PhishTank or PhishStats can result in the URL already being inactive by the time it's listed.
    \item Some phishing webpages are designed to self-destruct after a short period or after reaching a certain number of visits to avoid detection. This makes them inaccessible when revisited later.
    \item Phishing sites might block access based on geographic location or IP address. They may be operational but inaccessible from certain regions or networks.
    \item Phishers often abandon domains quickly and move to new ones to stay ahead of detection. As a result, many collected URLs might be from abandoned or moved domains.
\end{itemize}

Phish-Blitz fetches screenshots for $97.3\%$ of the legitimate URLs and $95.3\%$ of the phishing URLs provisioning to build large datasets for DL-based phishing webpage detection approaches.


\begin{table}[t]
    \centering
    \caption{Comparision of Phish-Blitz, \texttt{saveweb2zip}, \texttt{scrapio}}
    \label{tab:table}
    \begin{tabular}{llll}
    \toprule
    \textbf{Resource Components} & \textbf{Phish-Blitz} & \textbf{saveweb2zip} & \textbf{scrapio} \\ \midrule
    HTML & \ding{51} & \ding{51} & \ding{51} \\
    CSS & \ding{51} & \ding{51} & \ding{51} \\
    JS & \ding{51} & \ding{51} & \ding{51} \\
    images & \ding{51} & \ding{51} & \ding{51} \\
    favicon & \ding{51} & \ding{51} & \ding{51} \\
    links redirecting & \ding{51} & \ding{51} &  \\
    online screenshot & \ding{51} &  & \ding{51} \\
    offline screenshot & \ding{51} &  &  \\
    Screenshot comparision & \ding{51} & &  \\
    
    \bottomrule
    \end{tabular}
\end{table}

\subsection{Analysis}
\label{sec:Experimental_Analysis_with_Feature_Correlation}
For comparative analysis, we compare our tool Phish-Blitz with two of the pre-existing tools \texttt{saveweb2zip} and \texttt{scrapio} \cite{kulkarni2023bridging}. The difference in functionalities of these three tools is shown as a stack bar plot in Table~\ref{tab:table}. \texttt{saveweb2zip} tool can be used to fetch HTML, CSS, JS, image, favicon files for a URL. Only one URL can be given as input for this tool. \texttt{scrapio} tool can be used to fetch HTML, CSS, JS, image, favicon files along with the screenshot of the website. Phish-Blitz can be used to fetch HTML, CSS, JS, and image files along with two screenshots, one screenshot of the original webpage and another of the locally downloaded webpage (HTML file).

We perform a comparative analysis between Phish-Blitz and \texttt{saveweb2zip} to determine the efficiency of these two in terms of downloading webpage resources for a given set of URLs. In this study, we consider the same set of URLs as described in Section~\ref{sec:Experiment_Results}. The average time taken by Phish-Blitz and \texttt{saveweb2zip} to download the webpage resources for each URL are $18.21$ and $19.45$ seconds, respectively. Moreover, \texttt{saveweb2zip} does not capture webpage screenshots. But Phish-Blitz captures two webpage screenshots (original and local) and performs a comparative analysis between them, to verify whether the locally hosted webpage screenshot is visually similar to the original one. Phish-Blitz takes an average of $28.18$ seconds to capture both these screenshots. In addition, Phish-Blitz fetches resources from multiple URLs at once, but \texttt{saveweb2zip} can only fetch resources for one URL at a time. Phish-Blitz runs on localhost and \texttt{saveweb2zip} runs on web. Phish-Blitz downloads the webpage resources in a systematic directory structure, but \texttt{saveweb2zip} zips the directories and files and returns a zip file. The source code of our tool -- Phish-Blitz -- is publicly available on GitHub~\cite{our_tool}.



Phish-Blitz outperforms currently available tools for collecting web resources like \texttt{saveweb2zip} and \texttt{requests}. Phish-Blitz also has an edge on tools like \texttt{requests}, which are generally considered as bot by the websites and are not responded correctly by the server. Phish-Blitz has a mix of benefits as it uses \texttt{requests}, \texttt{selenium}, \texttt{wget}, \texttt{BeautifulSoup} together, making it very robust and efficient to use.

For the next step of comparative analysis, we compare Phish-Blitz with \texttt{saveweb2zip} and \texttt{scrapio} \cite{kulkarni2023bridging}. For this, we have taken screenshots as our medium for comparison. Also we have restricted our screenshot analysis to the legitimate URLs only, to avoid any discrepancies due to the short lifetime of phishing URLs. We have used the top 1000 legitimate URLs based on the PageRank. We have injected the screenshot code used in Phish-Blitz, into the code of \texttt{scrapio} \cite{kulkarni2023bridging} and to the code that fetches the web resources from \texttt{saveweb2zip}. For each URL, two screenshots are taken by each of these codes. The first screenshot would be of the online webpage of the URL provided. The second screenshot would be of the HTML file that has been downloaded by the code. These pictures are named \texttt{online.png} and \texttt{offline.png}. The \texttt{Selenium} module of Python is used to take the screenshot. The tool first fetches the height of the webpage and then use an inbuilt function in selenium called \texttt{save\_screenshot} to take the screenshot of the entire webpage. After fetching the screenshots, it checks the similarity in the two screenshots. For this part, an image comparative metric called Histogram Correlation\cite{histcorr} is used. OpenCV provides the \texttt{cv2.compareHist()} function to compare histograms, which offers various methods for comparison, including correlation. This correlation measures the linear relationship between two histograms, giving a score between $-1$ and $1$. The explanation about the scores is given in Table~\ref{tab:histcorr}.

\begin{table}[h]
    \centering
    \caption{Histogram Correlation}
    \label{tab:histcorr}
    \begin{tabular}{rl}
    \toprule
    \textbf{Score} & \textbf{Description} \\ \midrule
    \texttt{1} & Perfect correlation (the histograms are identical) \\
    \texttt{0} & No correlation \\
    \texttt{-1} & Inversely correlated (opposite distributions) \\

    \bottomrule
    \end{tabular}
\end{table}

Out of 1000 URLs, Phish-Blitz compares screenshots for 746 URLs, \texttt{saveweb2zip} compares 824 URLs and \texttt{scrapio} \cite{kulkarni2023bridging} compares 432 URLs. The reason for less number of comparisons as compared to \texttt{saveweb2zip} can be as follows:
\begin{itemize}
    \item Since Phish-Blitz runs on localhost, it relies on the local network and resources, which may have limitations in bandwidth or speed compared to the \texttt{saveweb2zip} which is a online tool that might be optimized with better infrastructure.
    \item Running locally may result in network or firewall restrictions blocking access to certain URLs. On a hosted service, these restrictions might not exist.
    \item Some websites may block requests from certain IP addresses, especially if they appear as local or residential IPs. The hosted service might use data centers or proxy IPs that are less likely to be blocked.
    
\end{itemize}

\begin{figure}[t]
    \centering
    \includegraphics[width=1\linewidth]{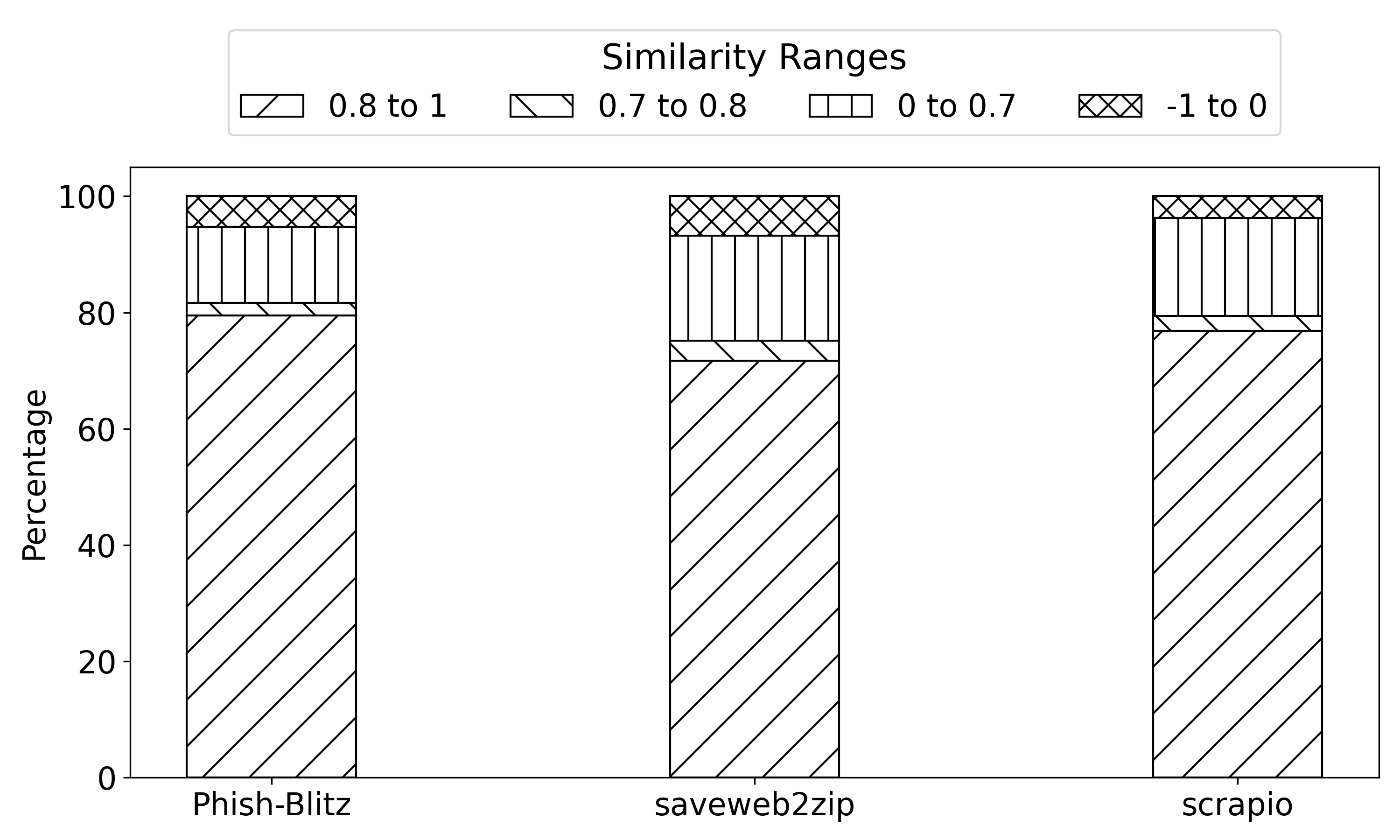}
    \caption{Screenshot comparison}
    \label{fig:comparsion}
\end{figure}

The comparative analysis of Phish-Blitz, \texttt{saveweb2zip} and \texttt{scrapio} \cite{kulkarni2023bridging} is shown in Figure \ref{fig:comparsion}.

Phish-Blitz can do more screenshot comparisons as compared to \texttt{scrapio} \cite{kulkarni2023bridging}. \texttt{saveweb2zip} can make more screenshot comparisons as compared to Phish-Blitz. But Phish-Blitz outperforms these two tools as Phish-Blitz can fetch more precise screenshots. Phish-Blitz successfully captures screenshots for $79.49\%$ of URLs, achieving over $80\%$ similarity to the original website. In comparison, \texttt{saveweb2zip} captures $71.72\%$ of URLs with the same accuracy threshold, while \texttt{scrapio} \cite{kulkarni2023bridging} captures $76.85\%$. This demonstrates that Phish-Blitz outperforms both alternatives in terms of maintaining high visual similarity to the original site. The huge similarity of the local file with the original website would imply that the tool is able to fetch all the required resources correctly and is able to successfully able to redirect the links to point to local resources. Table~\ref{tab:table} summarizes the differences between Phish-Blitz, \texttt{scrapio} \cite{kulkarni2023bridging} and \texttt{saveweb2zip}).



\section{Conclusion and Future Work}
\label{sec:Conclusion}
In this paper we developed a tool -- Phish-Blitz -- which can download all the required web resources i.e. HTML, CSS, JS, image files and screenshots, for legitimate and phishing URLs. The tool also redirects the \texttt{link} and \texttt{src} attributes of tags like \texttt{link}, \texttt{script}, and \texttt{img} to point to locally downloaded resources. This would help us to view the website offline without any discrepancy and also would help if these resources are taken down in future. \par The tool fetches legitimate URLs according to their PageRank. But these URLs do not get updated in real time according to their PageRank.
In future, we would like to add this feature which would fetch legitimate websites dynamically according to page rank. The tool currently uses PhishStats for fetching phishing URLs. We would like to extend this feature to various other sources also like OpenPhish, PhishStorm, PhishTank etc. The directory structure of downloaded web resources is highly dependent on the URL. We would like to make it URL independent in the future, as all the URLs will follow the same directory structure, which would help the user to browse all the web resources without any confusion. Some of the websites are not allowing the tool to take a screenshot, which is leading to taking screenshot of a blank page. We would like to improvise our code such that the tool is able to take screenshots of most of the URLs and also detect if the screenshot is a blank page or the screenshot of the website. Currently, Phish-Blitz concentrates on fetching the web resources only. In the future we would like to use this tool and develop a model for phishing webpage detection. 
\bibliographystyle{IEEEtran}
\balance
\bibliography{references}

\end{document}